\documentstyle[11pt,newpasp,twoside]{article}
\begin{document}
\title{Palomar 5 and its tidal tails: New observational results}
\author{M. Odenkirchen, E.K. Grebel, W. Dehnen, H.W. Rix}
\affil{MPI f\"ur Astronomie, K\"onigstuhl 17, D-69117 Heidelberg, Germany}
\author{C.M. Rockosi}
\affil{Univ. of Washington, Box 351580, Seattle, WA 98195-1580}
\author{H. Newberg}
\affil{Rensselaer Polytechnic Inst.,
110 Eighth Street, Troy, NY 12180-3590}
\author{B. Yanny} 
\affil{Fermilab, P.O.\ Box 500, Batavia, IL 60510}

\begin{abstract} 
Sloan Digital Sky Survey data for the field of the cluster Pal\,5 reveal 
the existence of a long massive stream of tidal debris spanning an 
arc of 10$^\circ$ on the sky. Pal\,5 thus provides an outstanding example 
for tidal disruption of globular clusters in the Milky Way.
Radial velocities from VLT spectra show that Pal\,5 has an extremely low 
velocity dispersion, in accordance with the very low mass derived from its 
total luminosity. 
\end{abstract}

\section{Extended tidal tails unveiled by the Sloan Survey}
Pal\,5 is a sparse low-mass halo cluster with peculiar structure
and stellar content. Using wide-field multicolor data from the 
Sloan Digital Sky Survey (SDSS, see York et al.\ 2000) we recently 
found clear direct 
evidence for strong mass loss from Pal\,5, showing that this cluster 
is in the process of being tidally disrupted (Odenkirchen et al.\ 2001, 
Rockosi et al.\ 2002). At the current stage the SDSS covers a 6$^\circ$ 
to 8$^\circ$ wide band across Pal\,5 (see Fig.1). 
This enabled us to extend our search for tidal debris to larger 
distances from the cluster. 
The data were filtered by applying an optimized smooth color-magnitude
dependent weight function. We thus found out that the tidal tails 
of Pal\,5 extend over at least 10$^\circ$ on the sky (Fig.1), 
corresponding to a length of 4 kpc in space. 
The leading tail (southwest of Pal\,5) is visible over 3\fdg5 down 
and most likely continues beyond the border of the field. 
The trailing tail (northeast of Pal,5) is traced out to 
6\fdg5 from the cluster. 
The stellar mass seen in the tails adds up to 1.2 times the 
mass of stars in the cluster. The location and curvature of the tails 
provide unique information on the local orbit of the cluster. 
The clumpiness of the stream suggests that the process of tidal mass
loss has been episodic, probably triggered by disk shocks. 
The orbit and the mass and geometry of the tails yield an
estimate of the mean mass loss rate of about 5~$M_\odot$/Myr.

\begin{figure}[h]
\plotfiddle{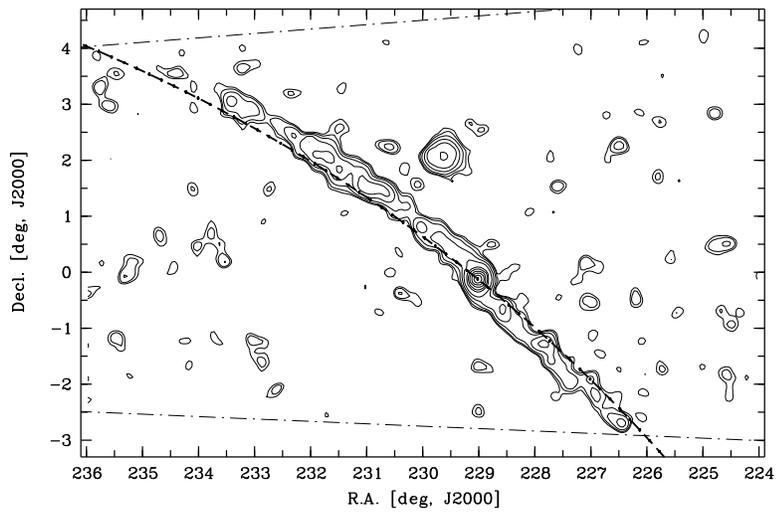}{6.8cm}{270}{50}{50}{-220}{235}
\caption{Map of the weighted stellar surface density showing the 
tails of Pal\,5 (contours drawn at $1.5 \sigma$, 
$2 \sigma$ and $3 \sigma$ and higher). The thick dashed line shows 
the best-fit orbit of the cluster. 
The feature at (230\fdg6,+2\fdg1) is due to the cluster M\,5 and 
hence not related to Pal\,5.
}
\end{figure}

\section{The velocity dispersion of the cluster}
In order to investigate the internal kinematics of Pal\,5 we 
obtained high resolution spectra of 18 of its red giants using 
the UVES spectrograph on the VLT (Odenkirchen et al.\ 2002). 
One of the stars has a velocity offset of 14 km\,s$^{-1}$ and is 
suspected to be a binary with rapid orbital motion. The others 
have highly coherent velocities, with an overall dispersion 
of 1.1~km\,s$^{-1}$. By accounting for the influence of binaries 
on the measured velocities in a statistical way we conclude that 
the dynamical line-of-sight velocity dispersion in Pal\,5 is 
probably in the range from 0.12 to 0.41~km\,s$^{-1}$. The 
upper end of this range is found to be compatible with the 
cluster's surface density profile and luminosity.\\


\vspace*{-1ex}

\acknowledgements
The SDSS is funded by the Alfred P. Sloan Foundation, the Participating 
Institutions, the National
Aeronautics and Space Administration, the National Science Foundation, 
the U.S. Department of Energy, the Japanese Monbukagakusho, and the
Max Planck Society. The SDSS web site is {\tt http://www.sdss.org/}.\\

\end{document}